\documentclass[12pt,aps,prb,preprint]{revtex4}   

\usepackage{amsmath,amssymb,amsfonts}
\usepackage[english]{babel}
\usepackage{graphicx}
\usepackage{dcolumn}
\usepackage{bm}

\begin{document}

\title{Interference with polarized light beams: Generation of spatially varying polarization}

\author{B. M. Rodr\'{\i}guez-Lara}
\email{bmlara@fisica.unam.mx}
\author{I. Ricardez-Vargas}
\email{ibisrv@fisica.unam.mx}
\affiliation{Instituto de F\'{\i}sica, Universidad Nacional Aut\'{o}noma de 
M\'{e}xico, Apdo. Postal 20-364, M\'{e}xico D.F. 01000, M\'{e}xico.}

\date{\today}

\begin{abstract}
Using a scheme based on a Mach-Zehnder interferometer, we propose an analysis of the superposition of polarized laser beams at a given angle. The focus of our study is the spatially varying polarization state of the resulting field, also known as a polarization grating, generated by this setup. Our proposal combines a theoretical description of the resulting field in terms of its Stokes parameters with an experimental demonstration of the existence of such a polarization grating due to the effects of polarization on beam interference experiments. 
\end{abstract}

\maketitle

\section{Introduction}

Understanding interference has been seminal in optics. More than two centuries ago, Young presented his Bakerian Lecture which contained an experimental demonstration of the general law of interference of light\cite{Young1804}. Fifteen years later, Fresnel and Arago studied the effect of the polarization state of light beams in the phenomena of interference\cite{Arago1819}. Thus, through interference, evidence of the transverse wave nature of light was brought forward. 

In the second half of the twentieth century there were many studies on the interference of polarized light for the undergraduate laboratory. Various interferometric methods were proposed to carry out such experiments using (i) a  Young double slit experiment, covering both slits with different polarizing filters\cite{Hunt1970, Pescetti1972, Mallick1973}; (ii) the polarizing properties of the ordinary and extraordinary axis in a double refracting crystal of calcite \cite{ Pontiggia1970, Ferguson1983} or a nematic liquid crystal\cite{Carr1991}; (iii)  a Ronchi grating as beam splitter placing polarizing filters afterwards the grating and analyzing at the conjugate plane of the grating through a focusing element \cite{Henry1981}; (iv) a Mach-Zehnder interferometer with polarizing filters at each arm \cite{Kanseri2008}. Just to list a few schemes.

In the same period of time, a sound mathematical description for the Young double slit experiment with polarized light was presented characterizing the resulting field by the Stokes parameters\cite{Collet1971}. Also, an alternate mathematical formulation and a proposal for using this scheme in image processing involving Wollaston prisms was discussed\cite{Andres1985}. Recently, this was used to understand the concept of a quantum eraser, where the polarization of two macroscopic fields is manipulated to loose or restore interference fringes \cite{Jordan2001}.

In all of the aforementioned studies, discussions focused on understanding the resulting intensity pattern, which is directly related to the Fresnel-Arago laws. To our knowledge, little has been said about the polarization state of the resulting field; it is possible to obtain a $\theta$-linearly polarized field from the superposition of two collinear fields with right and left circular polarization, with the given angle $\theta \in (-90, 90]$ degrees defined by the phase difference between the superposing fields\cite{Mellen1990}.

We believe the analysis of the superposition of two light fields could go beyond the study of Fresnel-Arago laws. Our motivation comes from the fact that a rich spatially dependent polarization structure, known as a polarization grating, arises from the superposition of two non-collinear polarized light beams with different polarizations. This polarization grating can be used to simplify the fabrication of three dimensional periodic microstructres eliminating complicated procedures to optimize contrast and polarization\cite{Cai2002}. It can be recorded in polarization sensitive materials, like bacteriorhodopsin, to perform phase shifting shearing interferometry without neither mechanical moving parts nor mounted gratings\cite{Garbusi2004}. It can also be used to simultaneously rotate or control the orientation of multiple microscopic birefringent particles, which might be useful for biomedical applications\cite{Mohanty2005}. These are just a few examples involving polarization gratings in modern research.

Our experimental proposal uses a slightly modified Mach-Zehnder interferometer to produce light with spatially varying polarization. Our theoretical description of the experiment is based on Jones calculus\cite{Jones1941} and Stokes parameters\cite{Collet1968} for polarized monochromatic light. The analysis is complemented with three specific examples involving the superposition of combinations of linearly and circularly polarized light showing some specific polarization gratings that can be produced with the proposed scheme. The presence of such polarization structures is experimentally confirmed through qualitative analysis of the resulting light field with a linear polarizer. We encourage the reader to reproduce the experiment and calculate the values for the Stokes parameters of the resulting field\cite{Gori1999, Schaefer2007}.

For readers who may not be familiar with Jones calculus and Stokes parameters we suggest the textbooks by Simmons and Guttmann\cite{Simmons1970} or by Born and Wolf\cite{Born} in addition to the articles mentioned above\cite{Jones1941,Collet1968,Collet1971}. For those interested in an advanced treatment, we recommend as a starting point the articles by Tervo {\em et. al}\cite{Tervo2003} and Roychowdhury and Wolf\cite{Roychowdhury2005}, as well as the introductory book by Wolf\cite{Wolf}. Their analyses deal with polarization and coherence degree of superposed arbitrary electromagnetic fields in three dimensional space. 

\section{Theoretical analysis}

\begin{figure}[ht!]
\begin{center}
\includegraphics[width= 0.45 \textwidth]{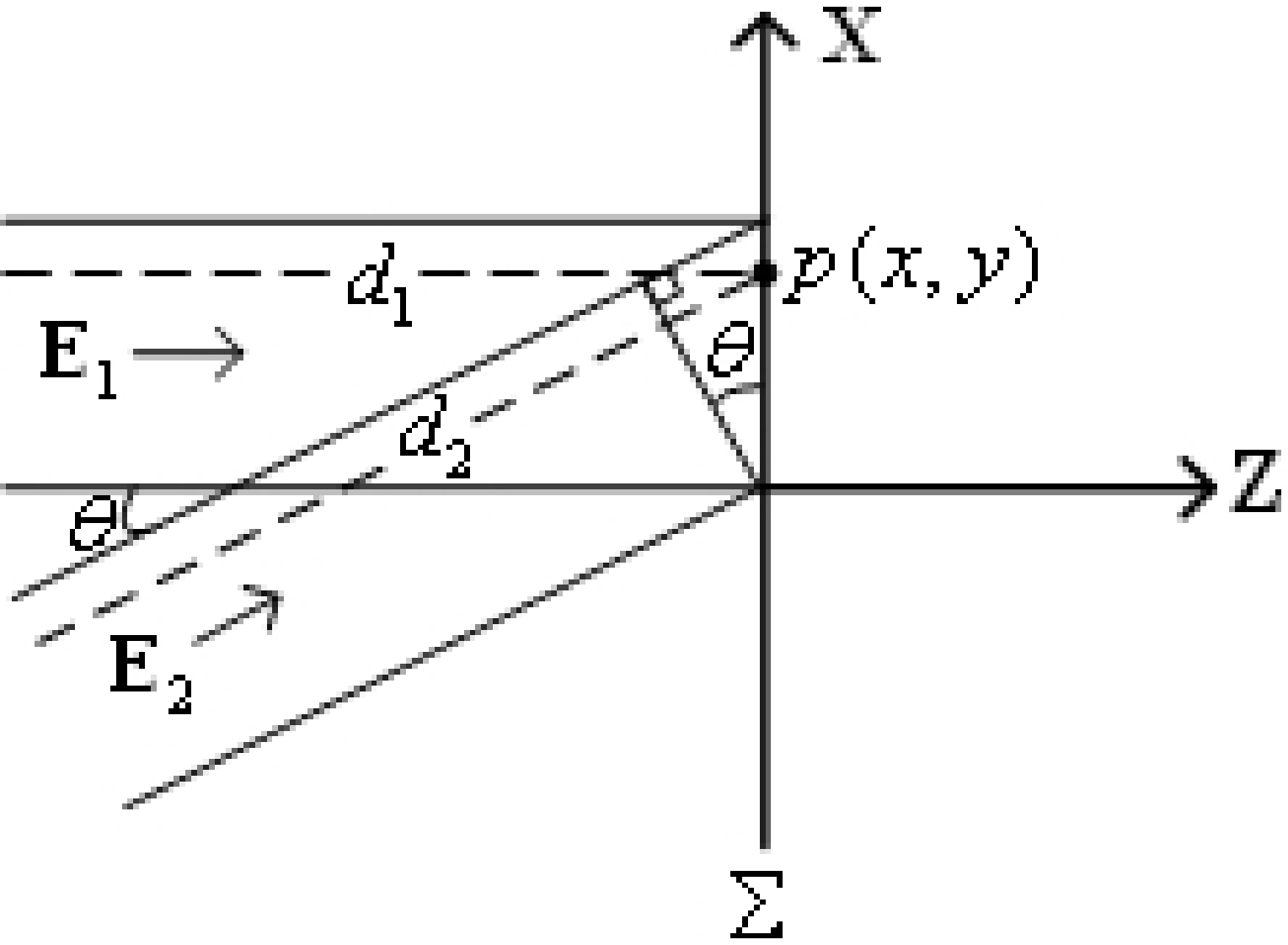}
\end{center}
\caption{\label{fig:Fig2} Theoretical simplification of the proposed experimental setup.}
\end{figure}

Figure \ref{fig:Fig2} illustrates a simplified version of the superposition scheme. We focus on the plane of incidence defined by the ${xz}$-plane so the $y$-coordinate will be obviated. Two polarized monochromatic plane waves of light intersect with a small angle $\theta$, such that $\sin \theta \approx \theta$ in radians, at some point $p(x,z)= x \hat{\bm{x}} + z \hat{\bm{z}}$ on the detection line $\Sigma$. Such light fields are described by the equations
\begin{eqnarray}
\bm{E}_{1}(x,z,t) &=& E_{1} ~e^{\imath (k d_{1}(x,z) - \omega t + \phi_{1})} \hat{\bm{\varepsilon}}_{1}(\alpha_{1},\delta_{1}), \nonumber \\
\bm{E}_{2}(x,z,t) &=& E_{2} ~e^{\imath (k d_{2}(x,z) - \omega t + \phi_{2})} R_{y}(\theta) \hat{\bm{\varepsilon}}_{2}(\alpha_{2},\delta_{2}) .
\end{eqnarray}
where the distances $d_{i}(x,z)$ are the distances from the $i$-th beam source to the point $p(x,z)$, e.g. $d_{1}(x,z)=z$. The counterclockwise rotation of the polarization state of the second beam about the $y$-axis is introduced  by means of the rotation matrix
\begin{equation}
R_{y}(\theta) = \left( \begin{array}{ccc} \cos \theta & 0 & \sin \theta \\ 0&1&0  \\ -\sin \theta & 0 & \cos \theta \end{array} \right).
\end{equation}
The unitary polarization vector state $\hat{\bm{\varepsilon}}_{j}(\alpha_{j},\delta_{j})$ is, up to a phase constant, a Jones vector
\begin{equation}
\hat{\bm{\varepsilon}}_{j}(\alpha_{j},\delta_{j}) = \cos \alpha_{j} ~\hat{\bm{x}}+ e^{\imath \delta_{j}} \sin \alpha_{j}~ \hat{\bm{y}}, 
\end{equation} 
with parameters in the ranges $\alpha_{j} \in [0, \pi/2]$ and $\delta_{j} \in (-\pi, \pi]$. The symbols $\hat{\bm{x}}$ and $\hat{\bm{y}}$ are the unitary vectors in the $x$- and $y$--directions.

The Stokes parameters for the total field $\bm{E}(x,z,t) = \bm{E}_{1}(x,z,t) + \bm{E}_{2}(x,z,t)$, at a point $p(x,z)$ on the detection line $\Sigma$ are given by the expression
\begin{eqnarray} \label{eq:StokesLarge}
S_{i} &=& \langle \bm{E}(x,z,t),\sigma_{i} \bm{E}(x,z,t) \rangle \nonumber \\
&=& s^{(1)}_{i} E_{1}^{2} + s^{(2)}_{i} E_{2}^{2} + 2 E_{1} E_{2} ~\text{Re} \left[ e^{\imath \Delta\Phi} \hat{\bm{\varepsilon}}_{1}^{\ast} \cdot \sigma_{i} R_{y}(\theta) \hat{\bm{\varepsilon}}_{2}  \right]. 
\end{eqnarray}
The angle brackets are shorthand notation for time averaging over the detection interval, which is large compared to the period associated with the optical radiation frequency,  $\langle {\bm{u}(r,t)}, {\bm{v}(r,t)} \rangle= \frac{1}{2} {\bm{u}(r)}^{\ast} \cdot {\bm{v}(r)}$ for plane waves (asterisk meaning complex conjugation). The symbol $\sigma_{i}$ for $i=0,1,2,3$ denotes the Pauli matrices
\begin{eqnarray}
\sigma_{0}= \left( \begin{array}{cc} 1&0\\0&1\end{array} \right), \mbox{ } 
\sigma_{1}= \left( \begin{array}{cc} 1&0\\0&-1\end{array} \right), \nonumber \\
\sigma_{2}= \left( \begin{array}{cc} 0&1\\1&0\end{array} \right), \mbox{ }
\sigma_{3}=\imath \left( \begin{array}{cc} 0&-1\\1&0\end{array} \right).
\end{eqnarray}
The parameters $s^{(j)}_{i}$ are the Stokes parameters for the $j$-th polarization vector $\hat{\bm{\varepsilon}}_{j}(\alpha_{j}, \delta_{j})$
\begin{eqnarray}
s_{0}^{(j)} &=& \hat{\bm{\varepsilon}}_{j}^{\ast} \cdot \sigma_{0} \hat{\bm{\varepsilon}}_{j} = 1, \nonumber \\
s_{1}^{(j)} &=& \hat{\bm{\varepsilon}}_{j}^{\ast} \cdot \sigma_{1} \hat{\bm{\varepsilon}}_{j} = \cos 2 \alpha_{j}, \nonumber  \\
s_{2}^{(j)} &=& \hat{\bm{\varepsilon}}_{j}^{\ast} \cdot \sigma_{2} \hat{\bm{\varepsilon}}_{j} = \sin 2 \alpha_{j} \cos \delta_{j}, \nonumber  \\
s_{3}^{(j)} &=& \hat{\bm{\varepsilon}}_{j}^{\ast} \cdot \sigma_{3} \hat{\bm{\varepsilon}}_{j} = \sin 2 \alpha_{j} \sin \delta_{j}.
\end{eqnarray}
Finally, the phase difference parameter $\Delta\Phi$ can be approximated as
\begin{eqnarray}
\Delta\Phi &=& k \left(d_{2}-d_{1}\right) + \Delta \phi \nonumber \\
&=& k~ x ~\sin \theta + \Delta \phi \nonumber \\
&\approx& k~ x~ \theta + \Delta \phi,
\end{eqnarray} 
with the initial phase difference between the sources of the beams given by $\Delta\phi = \phi_{2}-\phi_{1}$. In our experimental scheme the source for both beams is the same laser so the initial phase difference is null, $\Delta\phi = 0$. The experimental setup uses small angles, $\theta \le 10^{-3}$ radians, that for practical purposes, $\cos \theta \approx 1 - \theta^2$ and $\sin \theta \approx \theta$. These experimentally feasible restrictions allow us to consider the polarization state of the second field beam $\bm{E}_{2}$ in its own reference frame and in the general reference frame almost equal; i.~e. the Jones vectors  $\hat{\bm{\varepsilon}}_{2}(\alpha_{2},\delta_{2})$ and $R_{y}(\theta) \hat{\bm{\varepsilon}}_{2}(\alpha_{2},\delta_{2})$ are almost parallel,
\begin{eqnarray}
\langle \hat{\bm{\varepsilon}}_{2}(\alpha_{2},\delta_{2}), R_{y}(\theta) \hat{\bm{\varepsilon}}_{2}(\alpha_{2},\delta_{2}) \rangle &=&  \cos \theta +  \imath \sin \theta \sin 2 \alpha \sin \delta , \\ \nonumber
&\approx & 1 - \theta^2 + \imath \theta \sin 2 \alpha \sin \delta, \mbox{ } \theta \le 10^{-3}, \\ \nonumber
&\sim & 1.
\end{eqnarray}
These approximations simplify the theoretical treatment, thus the real parts involved in the last term of Eq.\eqref{eq:StokesLarge} are given by
\begin{eqnarray}
\text{Re} \left( e^{\imath \Delta\Phi}  \hat{\bm{\varepsilon}}_{1}^{\ast} \cdot  \sigma_{0} R_{y}(\theta) \hat{\bm{\varepsilon}}_{2}   \right) &\approx&  \cos \alpha_{1} \cos \alpha_{2} \cos \Delta\Phi + \sin \alpha_{1} \sin \alpha_{2} \cos \left(\Delta\Phi + \Delta \delta \right) , \nonumber \\
\text{Re} \left( e^{\imath \Delta\Phi}  \hat{\bm{\varepsilon}}_{1}^{\ast} \cdot  \sigma_{1} R_{y}(\theta) \hat{\bm{\varepsilon}}_{2}   \right) &\approx&  \cos \alpha_{1} \cos \alpha_{2} \cos\Delta\Phi - \sin \alpha_{1} \sin \alpha_{2} \cos \left( \Delta\Phi + \Delta \delta \right) , \nonumber \\
\text{Re} \left( e^{\imath \Delta\Phi}  \hat{\bm{\varepsilon}}_{1}^{\ast} \cdot  \sigma_{2} R_{y}(\theta) \hat{\bm{\varepsilon}}_{2}   \right) &\approx&  \cos \alpha_{1} \sin \alpha_{2} \cos \left( \Delta\Phi + \delta_{2} \right) + \sin \alpha_{1} \cos \alpha_{2} \cos \left( \Delta\Phi - \delta_{1} \right) , \nonumber \\
\text{Re} \left( e^{\imath \Delta\Phi}  \hat{\bm{\varepsilon}}_{1}^{\ast} \cdot  \sigma_{3} R_{y}(\theta) \hat{\bm{\varepsilon}}_{2}   \right) &\approx&  \cos \alpha_{1} \sin \alpha_{2} \sin \left( \Delta\Phi + \delta_{2} \right) - \sin \alpha_{1} \cos \alpha_{2} \sin \left( \Delta\Phi - \delta_{1} \right).  
\end{eqnarray}
As usual, the Stokes parameter $S_{0}$ is useful for discussing the intensity profile at the detection line such as discussed by Pescetti\cite{Pescetti1972} or Collet\cite{Collet1971}, while the latter three parameters, $S_{1}$ to $S_{3}$, relate to the polarization state of the field.

Our purpose is to understand the polarization properties of the total field. In order to do so, let us consider the interfering beams carrying orthogonal polarizations, that is $\langle \hat{\bm{\varepsilon}}_{1}(\alpha_{1},\delta_{1}), \hat{\bm{\varepsilon}}_{2}(\alpha_{2},\delta_{2}) \rangle = 0$. Notice that two orthogonal polarization vectors can be written as $\hat{\bm{\varepsilon}}_{1}(\alpha,\delta)$ and $\hat{\bm{\varepsilon}}_{2}(\alpha- \pi/2, \delta)$ if we relax the restrictions on the domain of $\alpha$. The Stokes parameters for orthogonal polarization vectors fulfill the condition $s_{i}^{(2)} = -s^{(1)}_{i}$ for $i=1,2,3$; the points on the polarization sphere that represent these vectors being antipodes. It is also possible to parametrize the amplitudes of the fields as $ \beta = \arctan \frac{ E_{2} }{ E_{1}}$ in a range $\beta \in [0,\pi/2]$, such that the corresponding normalized Stokes parameters for the total electromagnetic field on the detection line are
\begin{eqnarray}
\tilde{S}_{0} &\approx& 1, \nonumber \\
\tilde{S}_{1} &\approx& \cos 2 \alpha \cos 2 \beta + \sin 2 \alpha \sin 2 \beta \cos \Delta\Phi, \nonumber \\
\tilde{S}_{2} &\approx& \sin 2 \alpha \cos 2 \beta \cos \delta - \sin 2 \beta \left( \cos 2 \alpha  \cos \delta \cos \Delta\Phi  - \sin \delta \sin \Delta\Phi \right), \nonumber \\
\tilde{S}_{3} &\approx& \sin 2 \alpha \cos 2 \beta \sin \delta - \sin 2 \beta \left( \cos 2 \alpha  \sin \delta \cos \Delta\Phi  + \cos \delta \sin \Delta\Phi \right).
\end{eqnarray}
It is possible to write the latter three normalized Stokes parameters, $\tilde{S}_{1}$ to $\tilde{S}_{3}$, as the vector, 
\begin{eqnarray} 
\vec{\tilde{S}} &=& \tilde{S}_{1}~\hat{\bm s}_{1}+\tilde{S}_{2}~\hat{\bm s}_{2}+\tilde{S}_{3}~\hat{\bm s}_{3} \nonumber \\
&\approx& R_{s_{1}}(\pi - \delta) R_{s_{3}}(2 \alpha) \left( \sin 2 \beta ~ \vec{g} + \cos 2 \beta  ~ \hat{\bm s}_{1} \right),\label{eq:OrthogonalEq}
\end{eqnarray}
with the vector $\vec{g}$ defining a great circle on the $s_{2}s_{3}$-plane of the polarization sphere,
\begin{equation} 
\vec{g} = \cos \Delta\Phi~\hat{\bm s}_{2} +\sin \Delta\Phi~\hat{\bm s}_{3},
\end{equation} 
and the rotation matrices given in the traditional way
\begin{eqnarray}
R_{s_1}(\vartheta) &=& \left( \begin{array}{ccc} 1 & 0 & 0 \\ 0 & \cos \vartheta & \sin \vartheta \\ 0& -\sin \vartheta & \cos \vartheta \end{array} \right), \nonumber \\
R_{s_3}(\vartheta) &=& \left( \begin{array}{ccc} \cos \vartheta & \sin \vartheta & 0 \\-\sin \vartheta & \cos \vartheta&0 \\0&0&1 \end{array} \right). 
\end{eqnarray}

\begin{figure}[ht!]
\begin{center}
\includegraphics[width= 0.9 \textwidth]{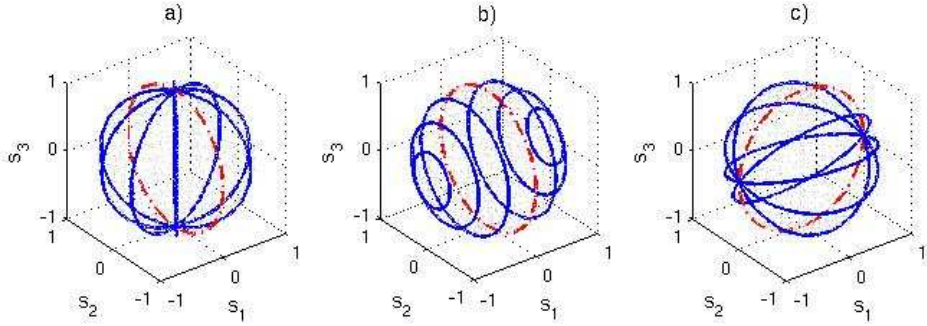}
\end{center}
\caption{\label{fig:parameters} (Color online) Effect of the parameters $\{\alpha, \beta, \delta\}$ for electromagnetic fields with orthogonal polarization, $\Delta\Phi \in [0,2 \pi)$. (a) Linear polarization $\alpha \in \{0, \pi/10, \pi/5, 3 \pi/10, 2 \pi/5 \}$, $\beta=\pi/4$, $\delta=0$; $\alpha=0$ in dot dashed red. (b) Linear polarization $\alpha = 0$, $\beta\in \{ \pi/16, \pi/8, 3 \pi/16, \pi /4, 5 \pi / 16, 3 \pi/8, 7 \pi/16 \}$, $\delta=0$; $\beta=\pi/4$ in dot dashed red.(c) Elliptical polarization $\alpha = \pi/4$, $\beta=\pi/4$, $\delta\in \{0, \pi/5, 2\pi/5, 3\pi/5, 4\pi/5\}$; $\delta=0$ in dot dashed red.}
\end{figure}

Equation \ref{eq:OrthogonalEq} implies that the parameter $\alpha$ generates a counterclockwise rotation around the $s_{3}$-axis, the parameter $\beta$ acts as a scaling factor and a $\hat{s}_{1}$-translation on the great circle $\vec{g}$, and the parameter $\delta$ as a counterclockwise rotation around the $s_{1}$-axis. \\
The counterclockwise rotations were expected. The great circle $\vec{g}$ is obtained from the superposition of the fields emitted by two sources with equal amplitudes of emission and horizontal/vertical linear polarizations.  The rotation $R_{s_{1}}(\pi - \delta) R_{s_{3}}(2 \alpha)$ transforms the two sets of Stokes parameters that map the orthogonal polarization pair $\hat{\bm{\varepsilon}}_{1}(0, 0)$, $\hat{\bm{\varepsilon}}_{2}(\pi/2, \pi)$ into those mapping any other orthogonal pair $\hat{\bm{\varepsilon}}_{1}(\alpha,\delta)$ and $\hat{\bm{\varepsilon}}_{2}(\alpha- \pi/2, \delta)$. Figure \ref{fig:parameters} shows an example of the effect of the set of parameters $\{ \alpha, \beta, \delta\}$ on the behavior of the polarization for the total field at the detection line. An interactive demonstration, where the user can input any combination of these parameters and obtain the corresponding Stokes parameters on the polarization sphere, is provided online\cite{WDP1}.

\section{Experimental setup and results}

We present the experimental realization and discussion of three cases that can shed more light on the problem when working in the undergrad laboratory. In the first two cases, beams are used with polarization states orthogonal to each other, Eq.\eqref{eq:OrthogonalEq}. In the third case, the general treatment is used, Eq.\eqref{eq:StokesLarge}. 

\begin{figure}[ht!]
\begin{center}
\includegraphics[width= 0.45 \textwidth]{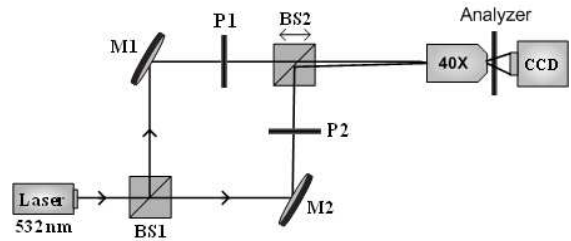}
\end{center}
\caption{\label{fig:ExpSetup} Experimental setup.}
\end{figure}

The experimental setup is shown in Fig. \ref{fig:ExpSetup}. Using a Mach-Zehnder interferometer, two beams are superposed at the back aperture of a microscope objective. The image of the beams superposition is formed at the focal region of the objective. Each of the beams is given a specific polarization state through a suitable retarder at the corresponding arm of the interferometer. Characterization of the superposition polarization state is performed by placing a linear polarizer as analyzer behind the focus of the objective. Images are captured for angles of $0$, $\pi/4$, $\pi/2$, and $3 \pi/4$ radians of the linear polarizer axis with respect to the horizontal axis.

A continuous wave solid state laser, emitting at a $532$~nm wavelength with a linear vertical output polarization state, ratio 100:1, is used as a source. The beam splitters, BS1 and BS2, are non-polarizing and one of them, BS2, is mounted on a linear displacement stage in order to control the angle of  interference, $\theta$ in Fig. \ref{fig:Fig2}. In the interferometer, each beam reflects once from a beam splitter and once from a mirror, thus canceling the phase shift introduced by each single reflection. A $40\times$ microscope objective is used as an imaging element. A black and white 1/2''-CCD, located just after the focal distance of the imaging element, is used to capture the images. 

The theoretical results shown here are calculated to fit the results at the CCD, whose range of detection is approximately $x \in [-6.5,6.5]$~mm. An intersection angle given by  $\theta \approx 6.1728 \times 10^{-4}$ radians is used.  

A general interactive demonstration where the user can set at will all of the two beams parameters is provided online\cite{WDP2}. This demonstration generates the corresponding Stokes parameters on the detection line, their representation on the polarization sphere, and the polarization ellipse for a given position on the detection line.

In the following paragraphs polarization states will be denoted using the notation: circularly right/left, $\mathcal{R}/\mathcal{L}$, elliptically right/left, $\mathcal{ER}/\mathcal{EL}$, linearly $\theta$, $\mathcal{P}(\theta)$.

\subsection{P-S Configuration. \\ Balanced horizontal/vertical linear polarization\\ 
$\alpha_1 = 0$, $\alpha_2 = \pi/2$ , $\beta=\pi/4$, $\delta_{1}=0$, $\delta_{2}=\pi$. }

\begin{figure}[ht!]
\begin{center}
\includegraphics[width= 0.7 \textwidth]{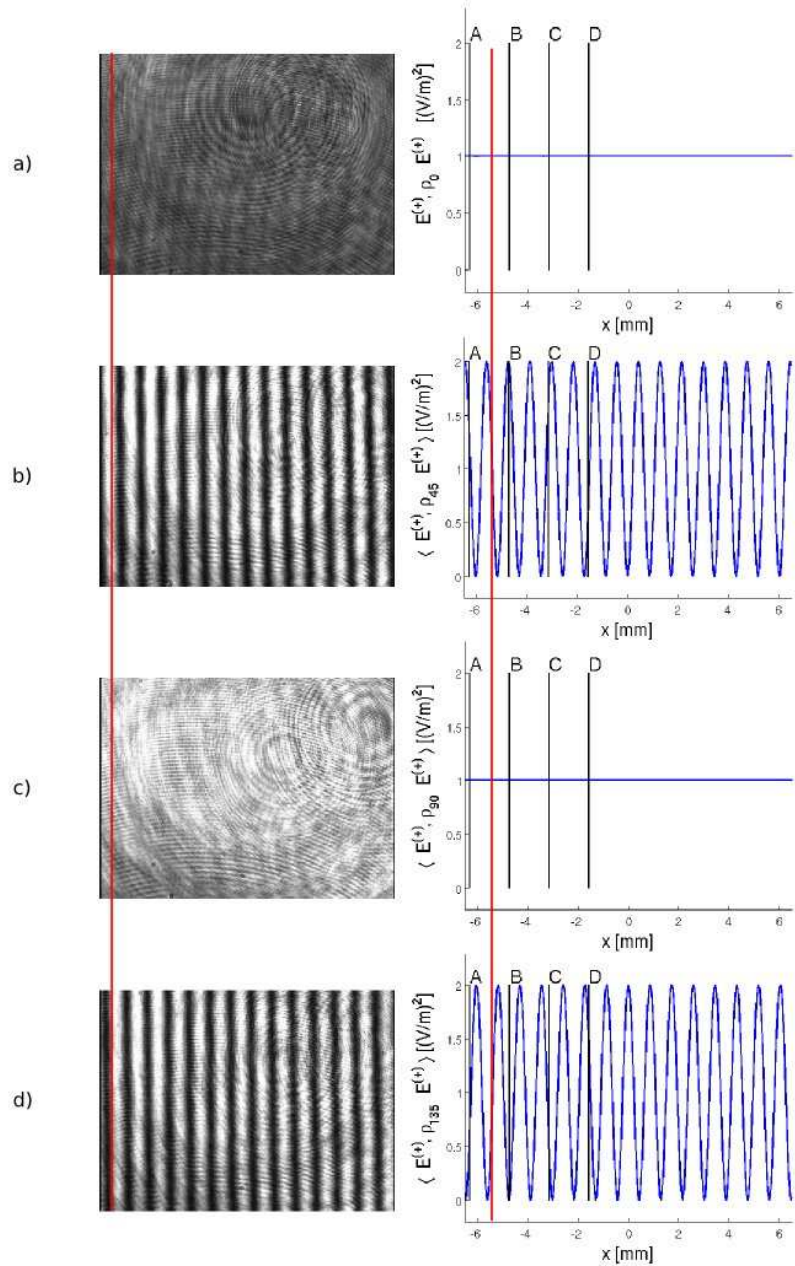}
\end{center}
\caption{\label{fig:case1a} (Color online) P-S Configuration. Interference of beams with horizontal and vertical linear polarization, equal field amplitudes and initial phases, $\alpha_{1}=0$ , $\alpha_{2}=\pi/2$, $\delta_{1}=0$, $\delta_{2}=\pi$, $\beta=\pi/4$. First column presents the experimental intensities obtained after the analyzer. Second column present the theoretical intensities. The orientation of the analyzer corresponds to (a) horizontal, (b) $45^{\circ}$, (c) vertical, (d) $-45^{\circ}$. The red vertical lines are presented as markers relating experimental and theoretical results.}
\end{figure}

In this case, P1 is a half-wave plate, whose fast axis is placed at $\pi/4$ radians with respect to the $x$-axis,  producing horizontal linear polarization and P2 is removed to keep vertical linear polarization. Figure \ref{fig:case1a} shows the intensities captured by the CCD camera at the four analyzer orientations for this configuration alongside the corresponding theoretical intensities. The vertical lines crossing both columns of figures are presented as reference marks to relate experimental with theoretical results. 

\begin{figure}[ht!]
\begin{center}
\includegraphics[width= 0.8 \textwidth]{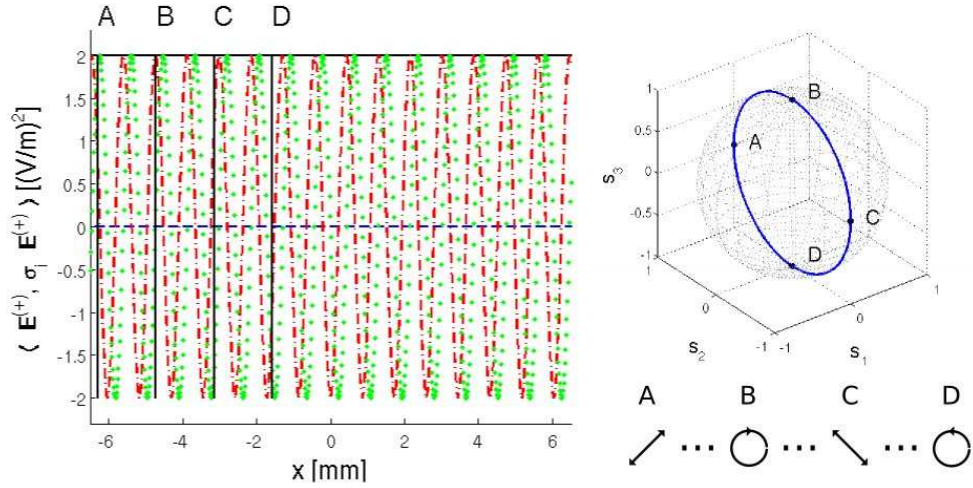}
\end{center}
\caption{\label{fig:case1b} (Color online) P-S Configuration. Interference of beams with horizontal and vertical linear polarization, equal field amplitudes and initial phases, $\alpha_{1}=0$ , $\alpha_{2}=\pi/2$, $\delta_{1}=0$, $\delta_{2}=\pi$, $\beta=\pi/4$. (Left) Stokes parameters $S_{0}$  (solid black), $S_{1}$  (dashed blue), $S_{2}$  (dot dashed red), $S_{3}$  (dotted green). (Top Right) Polarization trajectory on the polarization sphere given by the normalized Stokes parameters $\tilde{S}_{1}$, $\tilde{S}_{2}$, $\tilde{S}_{3}$. (Bottom Right) Polarization state corresponding to the point labeled A/B/C/D.}
\end{figure}

Figure \ref{fig:case1b} shows that the analysis is consistent with a polarization of the resulting field varying on a meridian of the polarization sphere crossing the $\pm 45^{\circ}$ linear polarization states; i.e. polarization varies periodically with the cycle: $\mathcal{R}$ $\rightarrow$ $\mathcal{P}(45)$ $\rightarrow$ $\mathcal{L}$ $\rightarrow$ $\mathcal{P}(-45^{\circ})$ $\rightarrow$  $\mathcal{R}$, with intermediate states of elliptical polarization. 

The alphabetical labels (A,B,C,D) in both figures identify the intensities of the analyzed polarization states in Fig. \ref{fig:case1a}, as well as the related Stokes parameters, their corresponding mapping on the polarization sphere, and the polarization state they represent in Fig. \ref{fig:case1b}, for the field at a given point $x$. Thus, the labels A/B/C/D show where the field has a $\mathcal{R}$/ $\mathcal{P}(-45^{\circ})$/$\mathcal{L}$/$\mathcal{P}(45^{\circ})$ polarization state.

\subsection{R-L Configuration. \\ Balanced right/left circular polarization\\ 
$\alpha_1 = \alpha_2 = \pi/4$ , $\beta=\pi/4$, $\delta_{1}= -\delta_{2}=\pi/2$. }
\begin{figure}[ht!]
\begin{center}
\includegraphics[width= 0.7 \textwidth]{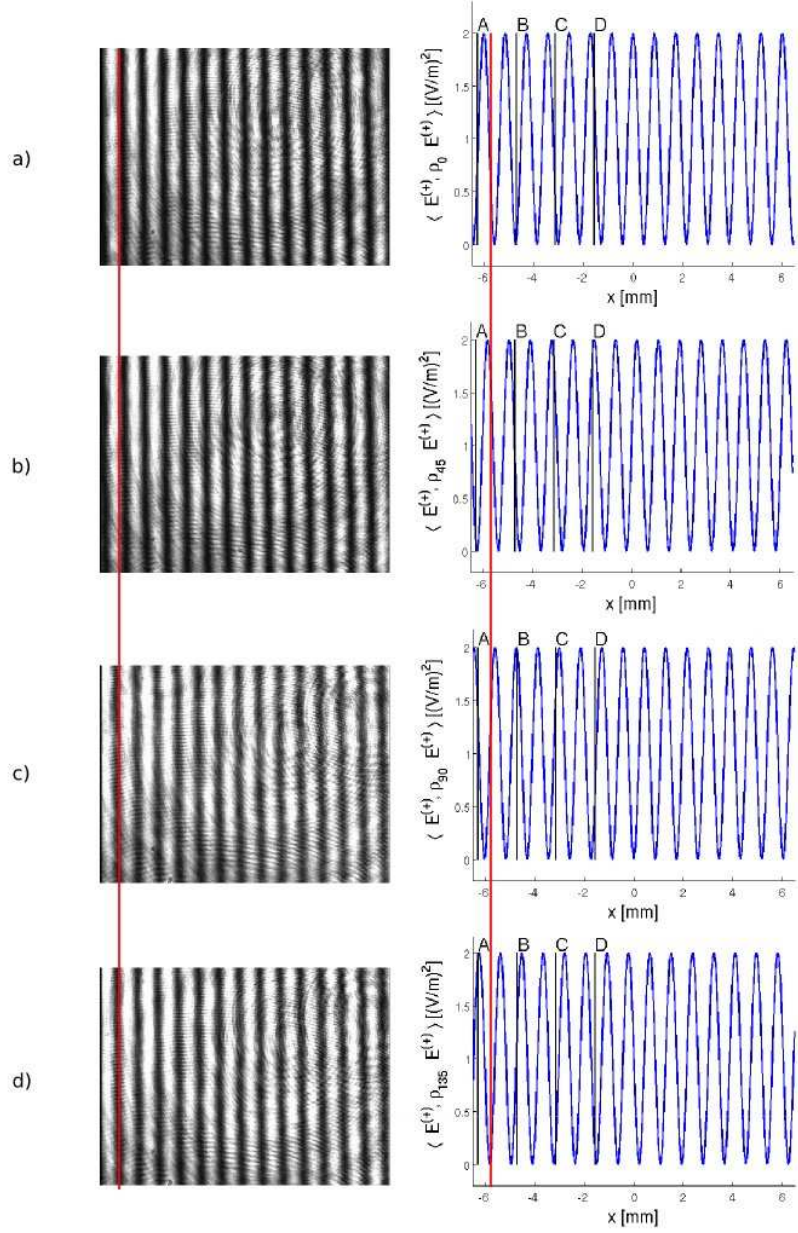}
\end{center}
\caption{\label{fig:case2a} (Color online) R-L Configuration.  Interference of beams with right 
circular and vertical linear polarization, equal field amplitudes and initial phases, $\alpha_{1}=\alpha_{2}=\pi/4$,  $\beta=\pi/4$, $\delta_{1}=-\delta_{2}=\pi/2$. First column presents the experimental intensities obtained after the analyzer. Second column present the theoretical intensities. The orientation of the analyzer corresponds to (a) horizontal, (b) $45^{\circ}$, (c) vertical, (d) $-45^{\circ}$. The red vertical lines are presented as markers relating experimental and theoretical results.}
\end{figure}

In this configuration, P1 and P2 are quarter-wave plates whose fast axes are placed at $\pm \pi/4$ radians with respect to the $x$-axis, thus their fast axes are perpendicular to each other,  producing right circular and left circular polarizations.  Figure \ref{fig:case2a} shows the experimental and theoretical intensities corresponding to the four analyzer positions. The vertical lines are used to relate both results.

\begin{figure}[ht!]
\begin{center}
\includegraphics[width= 0.8 \textwidth]{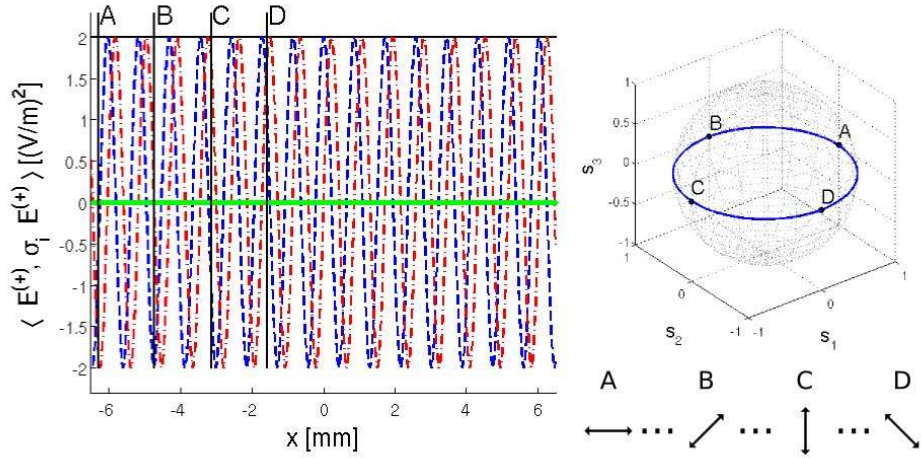}
\end{center}
\caption{\label{fig:case2b} (Color online) R-L Configuration. Interference of beams with right 
circular and vertical linear polarization, equal field amplitudes and initial phases, $\alpha_{1}=\alpha_{2}=\pi/4$,  $\beta=\pi/4$, $\delta_{1}=-\delta_{2}=\pi/2$.  (Left) Stokes parameters $S_{0}$  (solid black), $S_{1}$  (dashed blue), $S_{2}$  (dot dashed red), $S_{3}$  (dotted green). (Top Right) Polarization trajectory on the polarization sphere given by the normalized Stokes parameters $\tilde{S}_{1}$, $\tilde{S}_{2}$, $\tilde{S}_{3}$. (Bottom Right) Polarization state corresponding to the point labeled A/B/C/D.}
\end{figure}

Figure \ref{fig:case2b} shows that the analysis is consistent with a polarization of the resulting field varying on the equator of the polarization sphere; i.e. polarization is always linear with orientation angle varying periodically from $- \pi$ to $\pi$ radians. It is shown that ``The superposition of the right and left circularly polarized light yields linearly polarized light but the direction of the polarization depends on the phase angle between the two beams'' \cite{Mellen1990}. 

The alphabetical labels in both figures identify the intensities of the analyzed polarization states in Fig. \ref{fig:case2a}, as well as the related Stokes parameters, their corresponding mapping on the polarization sphere, and the polarization state they represent in Fig. \ref{fig:case2b}, for the field at a given point $x$. Thus, the label A/B/C/D shows where the field has a $\mathcal{P}(0^{\circ})$/ $\mathcal{P}(45^{\circ})$/$\mathcal{P}(90^{\circ})$/$\mathcal{P}(-45^{\circ})$ polarization state.

\subsection{R-S Configuration \\ Balanced horizontal linear and right circular polarization\\ 
$\alpha_1 = 0$, $\alpha_2 = \pi/4$, $\beta=\pi/4$, $\delta_{1}= 0$, $\delta_{2}=\pi/2$. }
\begin{figure}[ht!]
\begin{center}
\includegraphics[width= 0.7 \textwidth]{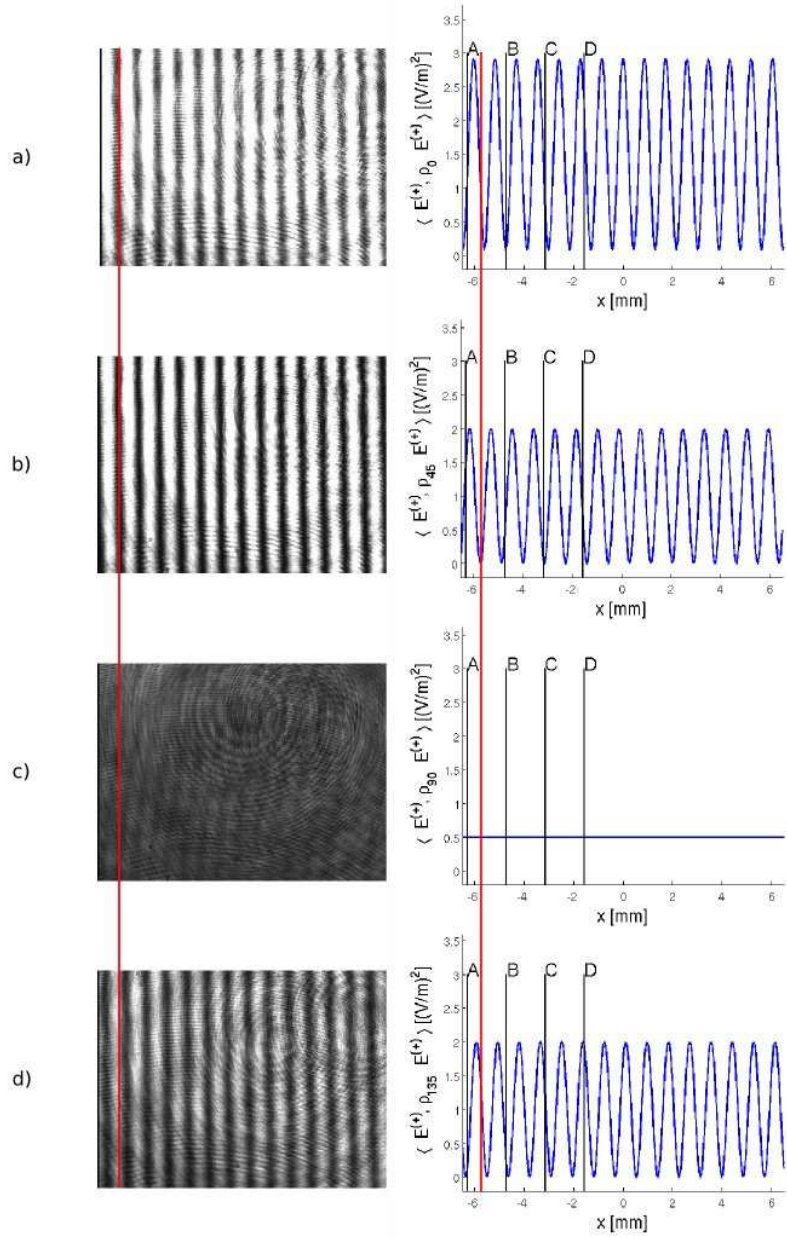}
\end{center}
\caption{\label{fig:case3a} (Color online) R-S Configuration.  Interference of beams with right and left
circular polarizations, equal field amplitudes and initial phases, $\alpha_{1}=0$ , $\alpha_{2}=\pi/4$, $\delta_{1}=0$, $\delta_{2}=\pi/2$, $E_{1}=E_{2}$. First column presents the experimental intensities obtained after the analyzer. Second column present the theoretical intensities. The orientation of the analyzer corresponds to (a) horizontal, (b) $45^{\circ}$, (c) vertical, (d) $-45^{\circ}$. The red vertical lines are presented as markers relating experimental and theoretical results.}
\end{figure}

Finally, here P1 is a quarter wave plate, whose fast axis is placed at $\pi/4$ radians with respect to the $x$-axis,  producing right circular polarization and P2 is a half-wave plate, whose fast axis is placed at $\pi/4$ radians with respect to the $x$-axis, producing horizontal linear polarization. Figure \ref{fig:case3a} shows the intensities captured by the CCD camera at the four analyzer orientations alongside the corresponding theoretical intensities. 

\begin{figure}[ht!]
\begin{center}
\includegraphics[width= 0.9 \textwidth]{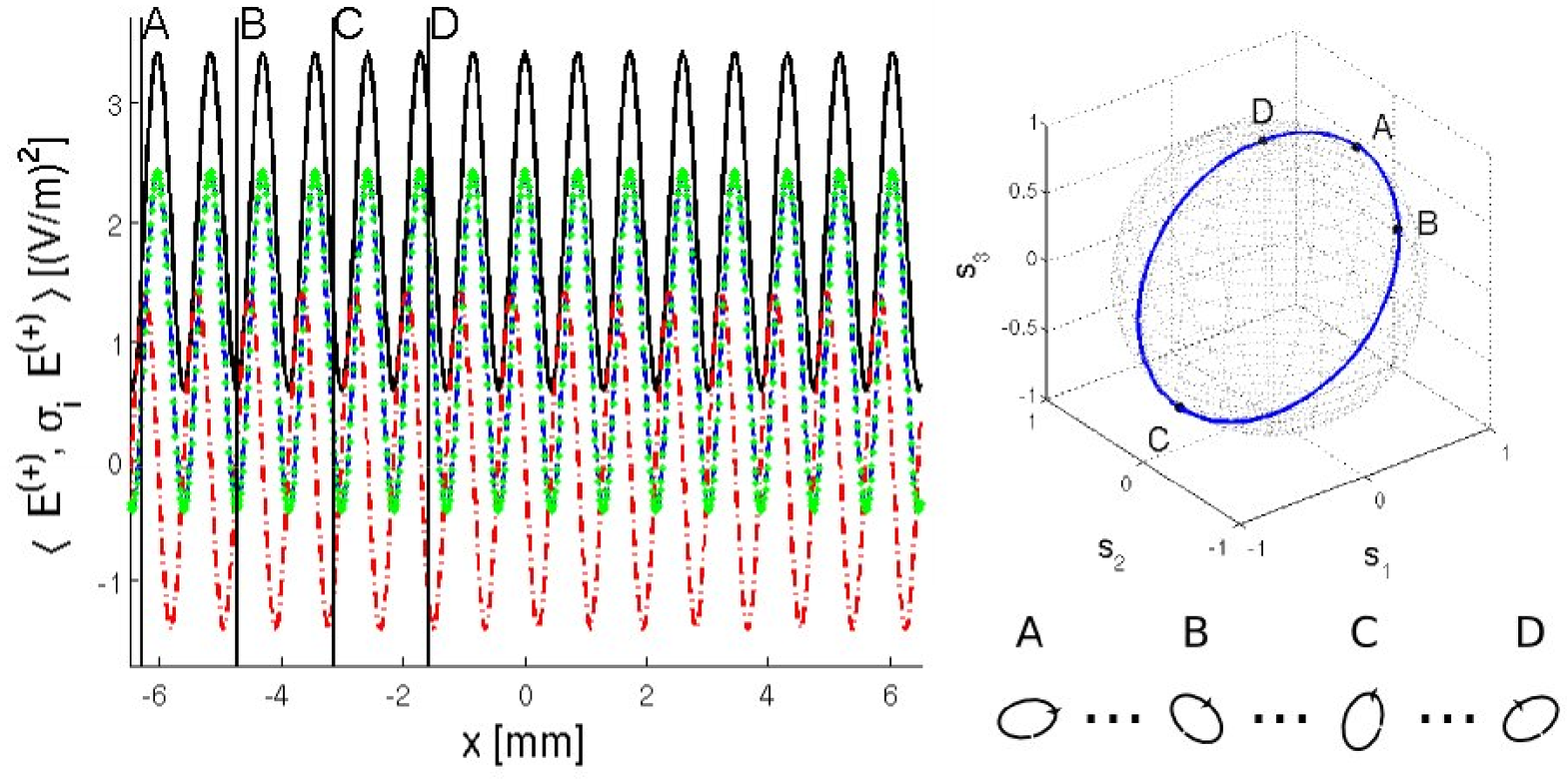}
\end{center}
\caption{\label{fig:case3b} (Color online) R-S Configuration. Interference of beams with right and left
circular polarizations, equal field amplitudes and initial phases, $\alpha_{1}=0$ , $\alpha_{2}=\pi/4$, $\delta_{1}=0$, $\delta_{2}=\pi/2$, $E_{1}=E_{2}$.  (Left) Stokes parameters $S_{0}$  (solid black), $S_{1}$  (dashed blue), $S_{2}$  (dot dashed red), $S_{3}$  (dotted green). (Top Right) Polarization trajectory on the polarization sphere given by the normalized Stokes parameters $\tilde{S}_{1}$, $\tilde{S}_{2}$, $\tilde{S}_{3}$. (Bottom Right) Polarization state corresponding to the point labeled A/B/C/D.}
\end{figure}

Figure \ref{fig:case3b} shows that the analysis is consistent with a polarization of the field varying on some circle on the polarization sphere; i.e. polarization varies periodically being elliptically polarized but for two points where it is linearly $\pm 45^{\circ}$ polarized. 

The alphabetical labels show elliptical polarization states with varying eccentricity and inclination of the major axis, for this configuration. Their intensities, Fig. \ref{fig:case3a} and their related Stokes parameters, their corresponding mapping on the polarization sphere, and the polarization state they represent, Fig. \ref{fig:case3b}, for the field at a given point $x$

\section{Conclusion}

We have presented an experimental scheme that an undergraduate student can use for analyzing the polarization state of the superposition of two slightly non-collinear polarized light beams. The equations modeling the Stokes parameters for this experiment have been presented. The explicit case of interfering orthogonal polarizations was discussed and complemented with two particular configurations to help elucidating this scheme; a third experimental configuration involving a general case, the interference of two non-orthogonal polarization beams, was also presented. It has been shown that the polarization state of light is spatially dependent in all cases due to the spatially dependent phase between the beams introduced by the impinging angle between them. 

Our experimental scheme can be implemented into an optical tweezer to demonstrate the transfer of intrinsic angular momentum to birefringent particles\cite{Moothoo2001} using polarization structures\cite{Mohanty2005}. This could also attract the attention to polarization, interference and mechanical properties of light at the undergraduate level.

\begin{acknowledgments}
B.~M.~R.~L. acknowledges financial support from UNAM-DGAPA. I.~R.~V. acknowledges financial support from CONACYT. Both authors thank R. J\'auregui and S. Hacyan for fruitful discussion. Authors are particularly grateful to K.~Volke-Sep\'{u}lveda.
\end{acknowledgments}

\newpage

\begin{thebibliography}{35}

\bibitem{Young1804} T. Young,``Experiments and calculations relative to physical optics'', Phil. Trans. R. Soc. Lond. {\bf 94}, 1--16 (1804).
\bibitem{Arago1819} D.~F.~J. Arago and A.~J.~Fresnel, ``On the action of rays of polarized light upon each other'', Ann. Chem. Phys. {\bf 2}, 288--304 (1819).

\bibitem{Hunt1970} J.~L. Hunt and G. Karl, ``Interference with polarized light beams'', Am. J. Phys. {\bf 38} 1249--1259 (1970). 
\bibitem{Pescetti1972} D. Pescetti, ``Interference of elliptically polarized light,'' Am. J. Phys. {\bf 40}, 735--740 (1972).
\bibitem{Mallick1973} S. Mallick, ``Interference with polarized light'', Am. J. Phys. {\bf 41}, 583--584 (1973)

\bibitem{Pontiggia1970} C. Pontiggia, ``Interference with polarized light'', Am. J. Phys. {\bf 39}, 679 (1971).
\bibitem{Ferguson1983} J.~L. Ferguson, ``A simple, bright demonstration of the interference of polarized light'', Am. J. Phys. {\bf 52}, 1141--1142 (1984).

\bibitem{Carr1991} E.~F. Carr and J.~P. McClymer, ``A laboratory experiment on interference of polarized light using a liquid crystal'', Am. J. Phys {\bf 59}, 366--367 (1991).

\bibitem{Henry1981} M. Henry, ``Fresnel-Arago laws for the interference in polarized light: A demonstration experiment'', Am. J. Phys. {\bf 49}, 690--691 (1981).

\bibitem{Kanseri2008} B. Kanseri, N.~S. Bisht, H.C. Kandpal, and S. Rath, ``Observation of the Fresnel and Arago laws using the Mach-Zehnder interferometer'', Am. J. Phys. {\bf 76}, 39--42 (2008).

\bibitem{Collet1971} E. Collet, ``Mathematical formulation of the interference laws of Fresnel and Arago '', Am. J. Phys. {\bf 39}, 1483--1495 (1971).

\bibitem{Andres1985} P. Andr\'es, A. Pons and J. Ojeda-Casta\~neda, ``Young's experiment with polarized light: Properties and applications'', Am. J. Phys. {\bf 53}, 1085--1088 (1985).

\bibitem{Jordan2001} T.~F. Jordan, ``Choosing and rechoosing to have or have not interference'', Am. J. Phys. {\bf 69}, 155--157 (2001).

\bibitem{Mellen1990} W.~R. Mellen, ``Interference patterns from circularly polarized light using a Michelson interferometer'', Am. J. Phys. {\bf 58}, 580--581 (1990).

\bibitem{Cai2002} L.~Z. Cai and X.~L. Yang, ``Interference of circularly polarized light: contrast and application in fabrication of three dimensional periodic microstructures'', Opt. Laser Technol. {\bf 34}, 671--674 (2002).

\bibitem{Garbusi2004} E. Garbusi, E.~M. Frins and J.~A. Ferrari, ``Phase-shifting shearing interferometry with a
variable polarization grating recorded on Bacteriorhodopsin'', Opt. Comm.  {\bf 241}, 309--314 (2004).

\bibitem{Mohanty2005} S.~K. Mohanty {\it et. al.}, ``Optical trap with spatially varying polarization'', Appl. Phys. B {\bf 80}, 631--534 (2005).

\bibitem{Jones1941} R.~C. Jones, ``A new calculus for the treatment of optical systems'', J. O. S. A. {\bf 31}, 488--493 (1941).

\bibitem{Collet1968} E. Collet, ``The description of polarization in classical physics'', Am. J. Phys. {\bf 36}, 713--725 (1968).

 
\bibitem{Gori1999} F. Gori, ``Measuring Stokes parameters by means of a polarization grating'', Opt. Lett. {\bf 24}, 584--586 (1999).
\bibitem{Schaefer2007} B. Schaefer {\it et. al.}, ``Measuring the Stokes polarization parameters'', Am. J. Phys. {\bf 75}, 163--169 (2007).

\bibitem{Simmons1970} J.~W. Simmons and Guttmann,  \textsl{States, waves and photons} (Addison--Wesley, Massachussetts, USA, 1970), 1th. ed. 

\bibitem{Born} M. Born and E. Wolf, \textsl{Principles of Optics} (Cambridge University Press, Cambridge, UK, 1999), 7th. ed.

\bibitem{Tervo2003} J. Tervo {\it et. al.}, ``Degree of coherence for electromagnetic fields'', Opt. Exp. {\bf 10}, 1137--1143 (2003).
\bibitem{Roychowdhury2005} H. Roychowdhury and E. Wolf, ``Young's interference experiment with light of any state of coherence and of polarization'', Opt. Comm. {\bf 252}, 268-274 (2005).

\bibitem{Wolf} E. Wolf,  \textsl{Introduction to the theory of coherence and polarization of light} (Cambridge University Press, Cambridge, UK, 2007), 1th. ed. 

\bibitem{WDP1} B.~M. Rodr\'{\i}guez-Lara and I. Ricardez-Vargas, ``Stokes Parameters for the Superposition of Two Slightly Noncollinear Light Beams with Orthogonal Polarizations'' from The Wolfram Demonstrations Project  http://demonstrations.wolfram.com/StokesParametersForTheSuperpositionOfTwoSlightlyNoncollinear/
 
\bibitem{WDP2} B.~M. Rodr\'{\i}guez-Lara and I. Ricardez-Vargas, ``Stokes Parameters for Superposition of Two Slightly Noncollinear Polarized Beams'' from The Wolfram Demonstrations Project http://demonstrations.wolfram.com/StokesParametersForSuperpositionOfTwoSlightlyNoncollinearPol/

\bibitem{Moothoo2001} D.~N. Moothoo {\it et. al.}, ``Beth's experiment using optical tweezers'', Am. J. Phys. {\bf 69}, 271--276 (2001).


\end{thebibliography}
\end{document}